\documentclass[aps,prl,twocolumn,amssymb,color,pdflatex]{revtex4}

\usepackage{graphicx}% Include figure files
\usepackage{graphics}
\usepackage{dcolumn}% Align table columns on decimal point
\usepackage{bm}% bold math
\usepackage{amsmath}
\begin{document}

\newcommand{\bk}{{\bf k}}
\newcommand{\bp}{{\bf p}}
\newcommand{\bv}{{\bf v}}
\newcommand{\bq}{{\bf q}}
\newcommand{\tbq}{\tilde{\bf q}}
\newcommand{\tq}{\tilde{q}}
\newcommand{\bQ}{{\bf Q}}
\newcommand{\br}{{\bf r}}
\newcommand{\bR}{{\bf R}}
\newcommand{\bB}{{\bf B}}
\newcommand{\bA}{{\bf A}}
\newcommand{\bE}{{\bf E}}
\newcommand{\bj}{{\bf j}}
\newcommand{\bK}{{\bf K}}
\newcommand{\cS}{{\cal S}}
\newcommand{\vd}{{v_\Delta}}
\newcommand{\tr}{{\rm Tr}}
\newcommand{\kslash}{\not\!k}
\newcommand{\qslash}{\not\!q}
\newcommand{\pslash}{\not\!p}
\newcommand{\rslash}{\not\!r}
\newcommand{\bs}{{\bar\sigma}}
\newcommand{\omt}{\tilde{\omega}}

\title{Origin of the T$_c$ enhancement in heterostructure cuprate superconductors}

\author{Doron L. Bergman and T. Pereg-Barnea}
\affiliation{Department of Physics,
California Institute of Technology, 1200 E. California Blvd, MC114-36,
Pasadena, CA 91125 }

\date{\today}

\begin{abstract}
Recent experiments on heterostructures composed of two or more films of cuprate superconductors of different oxygen doping levels\cite{Yuli,Gozar} have shown a remarkable T$_c$ enhancement (up to $50\%$) relative to single compound films.  We provide here a simple explanation of the enhancement which arises naturally from a collection of experimental works.  We show that the enhancement could be caused by a structural change in the lattice, namely an increase in the distance of the apical oxygen from the copper-oxygen plane.  This increase modifies the effective off-site interaction in the plane which in turn enhances the d-wave superconductivity order parameter.  To illustrate this point we study the extended Hubbard model using the fluctuation exchange approximation.
\end{abstract}
\maketitle
In 2008 Yuli {\it at al.}\cite{Yuli} presented a novel technique to enhance the superconducting transition temperature (T$_c$) of LSCO films.  The enhancement was achieved in a heterostructure of two thin layers of the same cuprate superconductor with different carrier density in the two layers.  The experimental setup used La$_{2-x}$Sr$_x$CuO$_4$ (LSCO) structures where the top layer had $x=0.35$ (overdoped) and in the bottom layer $x$ ranged from the underdoped to the overdoped.  In all measured heterostructures T$_c$ was higher in the heterostructure than that of a single compound LSCO film with the same doping $x$ as the bottom layer.  The largest enhancement occurred around $x=0.12$ where T$_c$ of the heterostructure was 32K, about 50\% larger than T$_c$ of the single compound film.

In another experiment Gozar {\it et al.}\cite{Gozar} created heterostructures with high T$_c$ from layers of metallic LSCO (overdoped) and either insulating LCO (the parent compound) or superconducting LCO (in this experiment the compound was LaCuO$_{4+\delta}$ with $\delta=0$ for the insulator).

These intriguing results present a possible new direction in the exploration of high temperature superconductors.  At the moment, the phenomenon of T$_c$ enhancement is far from understood and seems to depend crucially on materials and growth method\cite{Millo}.  A few ideas have been put forth regarding the mechanism by which the enhancement occurs.  These ideas are closely related to the different point of views on the origin of the pseudogap and its relation to the superconducting phase.

For example, if one takes the point of view that the pseudogap is related to an order parameter which competes with superconductivity then the tendency to develop such an order reduces T$_c$.  Therefore, if in the heterostructure this tendency is suppressed superconductivity will prevail to higher temperatures.  Some support for this point of view is provided by the experiment\cite{Yuli} in the following way.  The largest enhancement occurs close to 1/8 filling.  At this filling, the single compound films have a dip in the T$_c$ vs. $x$ curve.  The dip is believed to be due to charge stripes which are favorable at this filling. In the heterostructures no such dip is observed.

Another point of view of the pseudogap physics is that in the underdoped side of the superconducting dome, T$_c$ is restricted by phase fluctuations while pairing persists up to higher energy scale, possibly of the order of the pseudogap temperature, T$^*$.
Taking this point of view, it has been suggested by Yuli {\it et al.}\cite{Yuli} and theoretically explored by Berg {\it et al.}\cite{Berg} and by Goren and Altman\cite{Goren}, that in the heterostructure, enhancement may occur if phase fluctuations are suppressed compared to the single compound film.  To suppress the fluctuations the superfluid density should be increased, making vortices more costly in energy.  Higher superfluid density can be achieved by higher density of states at the Fermi level which can be provided by a metallic layer.  This view is supported by the fact that the enhancement only occurs on the underdoped side of the superconducting dome.

Both of the directions above are interesting and deserve further investigation.  In this work, however, we point at a different direction which does not necessarily discriminate between the two scenarios above but may lead the way to a microscopic description of the phenomenon.  We are guided by both experimental evidence and relevant numerical findings.  We first assume that the cause for the T$_c$ enhancement resides {\it within} the copper-oxygen plane.  This starting point should be contrasted with approaches that include a bilayer interface of the two materials and rely on strong hopping between the layers\cite{Berg,Goren}. Unfortunately, the
inter-layer hopping strength in the cuprate materials may be too weak to explain such strong variations in T$_c$. A quantitative measure of the interlayer hopping strength has been seen, for example, in transport properties\cite{Hosseini,Pereg-Barnea-lambda} and optical conductivity\cite{Homes}.  There is no evidence to suggest the heterostructures have stronger inter-plane coupling than that of the single compound film.

What is then the difference between the single compound film and the heterostructure?  One difference may be the electronic density.  The amount of doping $x$ in the bottom layer of the heterostructure may be different from that of the starting material due to charge migration between the two layers.  This leads to a 'self doping' effect which may change T$_c$.  This effect alone, however, cannot account for the large T$_c$ enhancement in the experiment.  Doping alone can not yield a transition temperature that is significantly higher than that of the single compound film at optimal doping.

Another difference is structural; atoms move and bonds stretch/shrink to relieve strain resulting from the two layer mismatch.  A recent experiment by Zhou {\it et al.}\cite{Zhou} provides evidence that structural changes are intimately related to the T$_c$ enhancement.  In these experiments heterostructures of the parent compound (LCO) and overdoped metallic LSCO are fabricated. These heterostructures also display superconductivity with an enhanced T$_c$ relative to the single film.  In order to detect {\it where} in the heterostructure superconductivity occurs, the experimental group introduce zinc impurities to the sample, layer by layer\cite{Logvenov}.  In-plane zinc impurities are known to suppress T$_c$ in the cuprates by a factor of  about 2.  In the heterostructure their effect was significant only when introduced to one specific layer which resides 1 layer above the interface, on the insulating/superconducting LCO side. This leads to the conclusion that only one layer is responsible for the heterostructure's high T$_c$.  In addition, the apical oxygen distance from each layer ($d$) was measured by x-ray scattering\cite{Zhou} and was found to vary significantly in the heterostructure, between $d \approx 2.3 \AA$ and
$d \approx 2.75 \AA$. In contrast, both the insulator (LCO) and metal (LSCO) have a bulk apical distance of $d \approx 2.4 \AA$  (as pointed out in Ref.~\onlinecite{Zhou} and measured in Ref.~\onlinecite{Radovic}). Therefore, the heterostructure achieves apical oxygen distances in doped LSCO, that are not reached in bulk samples.
We believe that this lattice rearrangement is responsible for the T$_c$ enhancement and motivate this point of view using a microscopic model below.

%The basic idea being that the heterostructure enjoys a more favorable apical oxygen distance for %SC, while not being limited by the doping of a bulk sample.

%In HgBa2CuO6 one finds cAÅ2.8 , longer by 0.9  than the in-plane Cu-O bond; coincidentally,
%this compound has the highest Tc ? 97 K among all single-CuO2-layer cuprates

In order to discuss superconductivity in the cuprates without speculating on the pairing mechanism and the origin of the pseudogap we explore the extended Hubbard model.  This model contains both the hopping parameters that are relevant to the cuprates and can be fit to the band structure as it is observed in experiment (e.g. ARPES\cite{Norman}) and the strongly correlated nature of this electronic system.  The Hamiltonian reads
\begin{eqnarray}\label{eq:Hubbard}
{\cal H} = -\sum_{ij,\sigma}t_{ij}c_{i\sigma}^\dagger c_{j\sigma} + U\sum_i n_{i\uparrow}n_{i\downarrow} + {1\over 2}V\sum_{\langle ij \rangle}n_i n_j
\end{eqnarray}
where the indices $i,j$ go over the lattice sites with angular brackets denoting nearest neighbors and $\sigma=\uparrow\downarrow$ is the spin label. The number of electrons at site $i$ is denoted by $n_i = n_{i\uparrow}+ n_{i\downarrow}= \sum_\sigma c_{i\sigma}^\dagger c_{i\sigma}$.

Numerical analysis of this model leads to d-wave superconductivity\cite{Onari}.  In order to see how this comes about one has to go beyond simple mean field and include strong spin and charge fluctuations. In this model the onsite repulsion $U$ promotes d-wave superconductivity while the off-site repulsion $V$ suppresses it.  This has been seen, for example, in the fluctuation exchange (FLEX) approximation\cite{Onari}.  In this approximation spin and charge fluctuations are taken into account (to all orders in perturbation theory) in the renormalized interaction vertex function.  This vertex together with the electronic Green's function is used in the Eliashberg\cite{Alexandrov} theory to determine the strength of pairing.  The analysis shows that when the off-site interaction $V$ is increased d-wave pairing is suppressed.  If $V$ is larger than some critical value a charge density wave becomes more energetically favorable than superconductivity.

The connection between the off-site interaction term and the apical oxygen distance from the Cu-O plane has been recently explored from first principles by Yin and Ku\cite{Yin} and by Weber {\it et al.}\cite{Weber}.  The conclusion of these studies is that the distance of the apical oxygen from the layer is inversely related to the bare $V$.  For materials with larger apical distance the off-site repulsion is smaller.  Other terms in the Hamiltonian, however, are rather insensitive to this distance.   The result is a d-wave superconductor with larger amplitude.  This is the main point of the paper - the apical oxygen distances in the heterostructure are significantly larger than in bulk LSCO, in particular at the superconducting layer. This causes a reduction of the off-site repulsion $V$ and superconductivity is therefore enhanced.

\begin{figure}[t]
\includegraphics[width = \columnwidth]{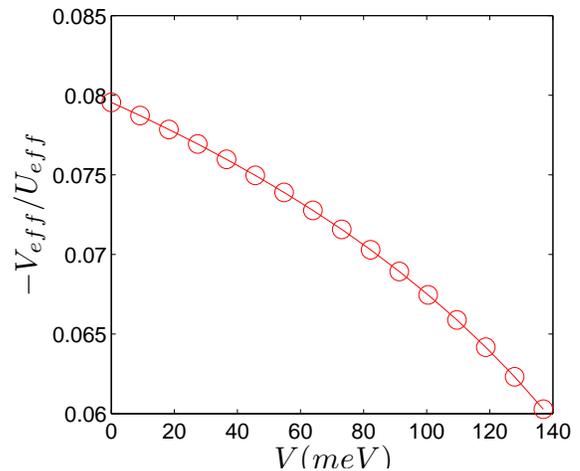}
\caption{The effective attractive off-site interaction $V_{eff}$ as a function of the bare repulsive, off-site, interaction $V$ in units of the on-site repulsion $U$. In the calculation the values $t = 488 meV$ and $U = 2.56 eV$ were used. The (bare) repulsive $V$ axis is in units of $meV$.
The values taken for the bare repulsive $V$ range from $V=0$ to its expected value for bulk LCO as found in Ref.~\onlinecite{Yin}.
}\label{fig:FLEX}
\end{figure}

In order to relate these findings to the experimental setup we investigate the Hamiltonian in Eq.~\ref{eq:Hubbard} as a model for the highest T$_c$ layer close to the interface.  We start with repulsive interactions both on-site ($U$) and off-site ($V$) and vary the strength of $V$ to capture the effect of the apical oxygen distance.  As a first step we use the FLEX approximation to include the Fermionic response in the effective interaction strength.  The renormalized interactions $U_{eff}$ and $V_{eff}$ are presented in Fig.~\ref{fig:FLEX}.  Note that the effective off-site interaction is {\it attractive}. Using renormalized interaction coefficients is crucial, since superconductivity requires pairing channel.

An intuitive understanding of the mechanism by which off-site interaction is related to superconductivity emerges from the FLEX approximation.  We start with a repulsive on-site and off-site interaction in the extended Hubbard model, as appropriate for this system.  However, when considering both interactions together in the vertex function the total off-site channel is attractive.  This can be seen even at the level of the standard Hubbard model (without off-site interaction).  The Hubbard model close to half filling was studied by Scalapino in Ref.~\cite{Scalapino} where it was found that the on-site repulsive interaction $U$ leads to an effective off-site attraction.  This is the result of proximity to the antiferromagnetic phase at half filling which is due to double hopping processes of the order of $t^2/U$.  Away from half filling there is no long range Ne\'el order but the susceptibility is still peaked around $(\pi,\pi)$.  The contribution of such a structure to the interaction vertex is, again, strongest at this point in momentum space which leads to a large attraction on nearest neighbor sites.  This can be viewed as large Friedel oscillations.  Electrons on nearest neighbor sites take advantage of these oscillations through pairing in the d-wave channel.

\begin{figure}[t]
\includegraphics[width = \columnwidth]{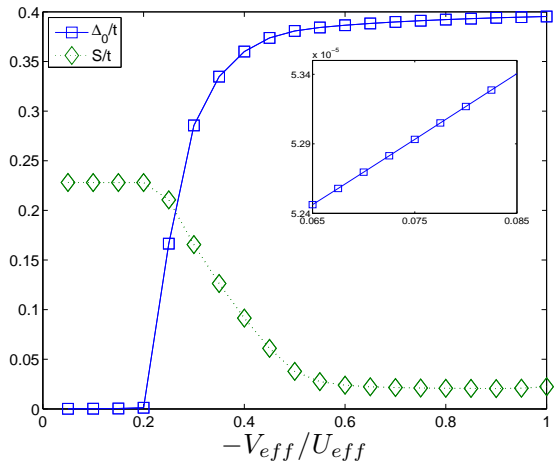}
\caption{The mean field result for d-wave superconductivity order parameter (squares) and Ne\'el order (diamonds) as a function of the ratio between the off-site interaction $V$ and the on-site interaction $U$.  The inset shows a zoom-in on the relevant interaction ratio for the superconducting order parameter $\Delta_0$.}\label{fig:MF}
\end{figure}

In order to determine T$_c$ realistically one should use the renormalized vertex in a full Eliashberg calculation as was done by Onari {\it et al.}\cite{Onari} In the present work we calculate effective interaction parameters
using the FLEX approximation, and turn to mean field theory in order to sketch the qualitative effect of the interaction on the order parameter.  
In the effective model the on-site interaction $U_{eff}$ is repulsive while the off-site interaction $V_{eff}$ is attractive.  The bare repulsion, and the induced effective attraction combine to give an effective {\it attractive} off-site interaction and a renormalized on-site repulsion.
When the apical oxygen distance is increased, the bare repulsion is decreased, and the overall effective interaction is more strongly attractive.  We show the mean-field result for the above effective Hubbard model in Fig~\ref{fig:MF}.  In this figure, as the ratio between the magnitude of the attractive $V_{eff}$ and the on-site interaction $U_{eff}$ is increased d-wave superconductivity is enhanced and the Ne\'el order (or spin density wave) is suppressed.  This result is obtained by the usual mean field decomposition of the interaction terms in the superconducting and antiferromagnetic channels.
The order parameters are found by minimizing the mean-field free energy.
%\begin{eqnarray}
%{1\over N}\sum_{\bk}{\chi_\bk^2 \over \sqrt{\epsilon_\bk^2 + \Delta_\bk^2+S^2}} = {8\over |V_{eff}|} \\
%{1\over N}\sum_{\bk}{1 \over \sqrt{\epsilon_\bk^2 + \Delta_\bk^2+S^2}} = {4\over U_{eff}}
%\end{eqnarray}
%\begin{widetext}
\begin{eqnarray}
{\cal F}&=& -\sum_{\bk \in RBZ}[E_1 + E_2]-NU(\bar{n}^2-S^2)-NV\Delta_0^2 \nonumber \\
E_1 &=& \sqrt{(US)^2 + (V\Delta_\bk)^2 +{\epsilon_\bk^2 + \epsilon_{\bk+\bQ}^2 \over 2}+(\epsilon_\bk + \epsilon_{\bk+\bQ})\xi} \nonumber\\
E_2 &=& \sqrt{(US)^2 + (V\Delta_\bk)^2 +{\epsilon_\bk^2 + \epsilon_{\bk+\bQ}^2 \over 2}-(\epsilon_\bk + \epsilon_{\bk+\bQ})\xi}\nonumber\\
\xi &=&\sqrt{(US)^2+\left({\epsilon_\bk-\epsilon_{\bk+\bQ} \over 2}\right)^2}
\end{eqnarray}
%\end{widetext}
where $RBZ$ stands for the sum over the reduced Brillouin zone, $\epsilon_\bk = -2t(\cos(k_x)+\cos(k_y))-\mu+U\bar{n}$ is the band dispersion which includes a chemical potential shift $U\bar{n}$, $\Delta_\bk = \Delta_0 \chi_\bk$ with $\chi_\bk = 2(\cos(k_x)-\cos(k_y))$ is the gap function, $S$ is the Ne\'el order parameter which folds the Brillouin zone with the wavevector $\bQ = (\pi,\pi)$ and $N$ is the number of sites.  We have used an attractive potential to derive these equations and write $|V_{eff}|$ to emphasize this point.
\begin{eqnarray} &
{U\over 2N}\sum_{\bk\in RBZ}\left[{1+(\epsilon_\bk+\epsilon_{\bk+\bQ})/2\xi\over E_1}+ {1-(\epsilon_\bk+\epsilon_{\bk+\bQ})/2\xi\over E_2}\right]=1 \nonumber \\ &
-{V\over 2N}\sum_{\bk\in RBZ}\chi_\bk^2\left[{1\over E_1}+ {1\over E_2}\right]=1 \nonumber \\ &
-{1\over 2N}\sum_{\bk\in RBZ}\left[{{\epsilon_\bk+\epsilon_{\bk+\bQ} \over2} +\xi \over E_1}+{{\epsilon_\bk+\epsilon_{\bk+\bQ} \over2} -\xi \over E_2}\right] = \bar{n} \nonumber\\
\end{eqnarray}
where the third equation has only a trivial solution ($\bar{n}=0$) at half filling.
%Please note that this crude mean field approximation over estimates the magnitude of the Ne\'el order parameter.
%%%%%%%% new
We must point out the numbers we find from our simple mean field treatment in Fig.~\ref{fig:MF}, suggest the superconductor is significant only if $\frac{|V_{eff}|}{U_{eff}} >  0.2 $, while Fig.~\ref{fig:FLEX} shows a value
no higher than $0.08$ for this ratio. However, in a full Eliashberg treatment even a small attractive off-site potential (like that of Fig.~\ref{fig:FLEX}) can cause the superconducting T$_c$ to be non-zero\cite{Onari}. We therefore conclude that our simplistic analysis gives the correct qualitative results, while leaving something to be desired in terms of quantitative results.

Before our conclusions, let us briefly discuss the possibility that phase fluctuations are suppressing T$_c$ on the underdoped side.  In the models above only the amplitude of the d-wave pairing order parameter is considered.  However, there is a lot of experimental evidence that phase fluctuations are important in the pseudogap phase.  Naturally, the question arises whether phase fluctuations are reducing T$_c$ from the temperature in which pairs are formed (which should be associated with a higher energy scale, possibly as high as T$^*$) or pairing and phase coherence appear at the same temperature (T$_c$) while the higher energy pseudogap is due to a competing order parameter. To shed more light on this issue it would be helpful to study the pseudogap regime of the heterostructure and measure T$^*$.
We propose to perform the following experiments.  Though experimentally challenging, it may be possible to measure the density of states (DOS) in the heterostructure and determine its pseudogap temperature T$^*$.  If we adopt the point of view that T$^*$ is due to a competing order, unrelated to superconductivity we expect this temperature to be lower in the heterostructure than in the single compound films.  This will reflect the competition; when the competing order is suppressed by a larger apical oxygen distance, d-wave superconductivity is enhanced.  If, on the other hand T$^*$ is the temperature at which pairing begins it should be higher in the heterostructure since T$_c$ and T$^*$ are related.  Another important measurement is that of the Nernst effect which has provided evidence for phase fluctuations in the past\cite{Ong} together with a diamagnetism probe\cite{Li}.  It is important to determine whether the phase fluctuations are enhanced or suppressed in the heterostructure relative to the single compound film.

In addition, we would like to propose a third experiment which would directly address the role of the apical oxygen in the T$_c$ enhancement.  Applying tensile stress along the c-axis will act to reduce the apical oxygen distance.  This, according to the scenario we present here should increase the bare off-site repulsion and reduce T$_c$.

To summarize, we provide a simple explanation to the enhancement of the superconducting transition temperature, T$_c$, in LSCO heterostructures compared with heterogenous, single compound films, as seen in recent experiments\cite{Yuli,Gozar}.
We have analyzed the extended Hubbard model in the FLEX approximation and used a heuristic mean-field analysis
to asses the effect of the apical oxygen distance on order formation. We conclude that the off-site nearest neighbor interaction has a tremendous effect on T$_c$ in d-wave superconductors.  In particular, when this interaction becomes more negative (in other words when it's bare repulsive component is reduced) the transition temperature increases.  This interaction has been found to be inversely related to the apical oxygen distance from the copper oxygen planes\cite{Yin,Weber}.  It has also been seen that the apical oxygen distance is larger
in the heterostructure, compared to the single compound film\cite{Zhou}.  We conclude that it is this structural difference between the heterostructure and the single compound film that is responsible for the enhancement of T$_c$.

The authors wish to acknowledge useful discussions with O. Millo, N. Lindner, G. Refael and N.-C. Yeh.
TPB acknowledges funding from the Research Corporation Cottrell Fellowship and DLB was supported by the Sherman Fairchild Foundation.

\bibliographystyle{apsrev}

%\bibliography{enhancement}

\end{document}